\documentclass[pre,reprint,showpacs]{revtex4-1}
\usepackage{graphicx}
\usepackage{amsmath,amssymb}
\usepackage{mhchem}

\usepackage{ulem}
\usepackage{color}

\newcommand{\me}{{\rm e}}
\newcommand{\md}{{\rm d}}

\begin{document}

\title{
Sequence selection by dynamical symmetry breaking in an
autocatalytic binary polymer model}

\author{Harold Fellermann}
\email{harold.fellermann@newcastle.ac.uk}
\affiliation{Interdisciplinary Computing and Complex Biosystem Research Group, School of Computing Science, Newcastle University, Claremont Tower, Newcastle upon Tyne NE1 7RU, United Kingdom}

\author{Shinpei Tanaka}
\affiliation{Graduate School of Integrated Arts and Sciences, Hiroshima
University, 1-7-1 Kagamiyama, Higashi-Hiroshima 739-8521, Japan}

\author{Steen Rasmussen}
\affiliation{Center for Fundamental Living Technology (FLinT) Department of Physics, Chemistry and Pharmacy, University of Southern Denmark, Campusvej 55, 5230 Odense M, Denmark}
\affiliation{Santa Fe Institute, 1399 Hyde Park Rd, Santa Fe NM 87501, USA}

\pacs{
      05.65.+b 
      87.23.Cc 
      87.23.Kg 
      82.40.Qt 
}

\begin{abstract}
Template directed replication of nucleic acids is at the essence of all living
beings and a major milestone for any origin of life scenario.
We here present an idealized model of prebiotic sequence replication, where
binary polymers act as templates for their autocatalytic replication, thereby
serving as each others reactants and products in an intertwined molecular ecology.
Our model demonstrates how autocatalysis alters the qualitative and quantitative
system dynamics in counter-intuitive ways.
Most notably, numerical simulations reveal a very strong intrinsic selection
mechanism that favors the appearance of a few population structures with highly
ordered and repetitive sequence patterns when starting from a pool of monomers.
We demonstrate both analytically and through simulation how this ``selection of
the dullest'' is caused by continued symmetry breaking through random
fluctuations in the transient dynamics that are amplified by autocatalysis and
eventually propagate to the population level.
The impact of these observations on related prebiotic mathematical models is discussed.
\end{abstract}
\maketitle 
%

\section{Introduction}
The ability of nucleic acids to serve as templates for their own replication via
the well-known Watson-Crick pairing is at the heart of all life known today.
Consequently, the onset of such chemical sequence information and its
replication are crucial cornerstones in most theories about the origin of
life~\cite{eigen_selforganization_1971,gilbert_rna_1986}.


While the appearance of first molecular replicators is still subject to ongoing
debate~\cite{wong_dawn_2008,powner_synthesis_2009}, we
suspect~\cite{declue_nucleobase_2009,rasmussen_generating_2016} that catalysis
combined with replication of the catalytic molecules is the critical molecular
invention.
Catalysis enables access to energy and resources while replication both
preserves and  enables combinatorial exploration of the catalyst.  
Competition over common and scarce resources would cause a selection pressure on
the catalyst that directs the random exploration process of copy error
mutations.
The resulting evolutionary search would eventually select replicator species
with increasingly  advantageous catalytic
properties~\cite{eigen_transfer-rna:_1981} -- thus exemplifying Darwinian
chemical evolution.

\begin{figure}[h!]
\begin{center}
\includegraphics[width=\columnwidth]{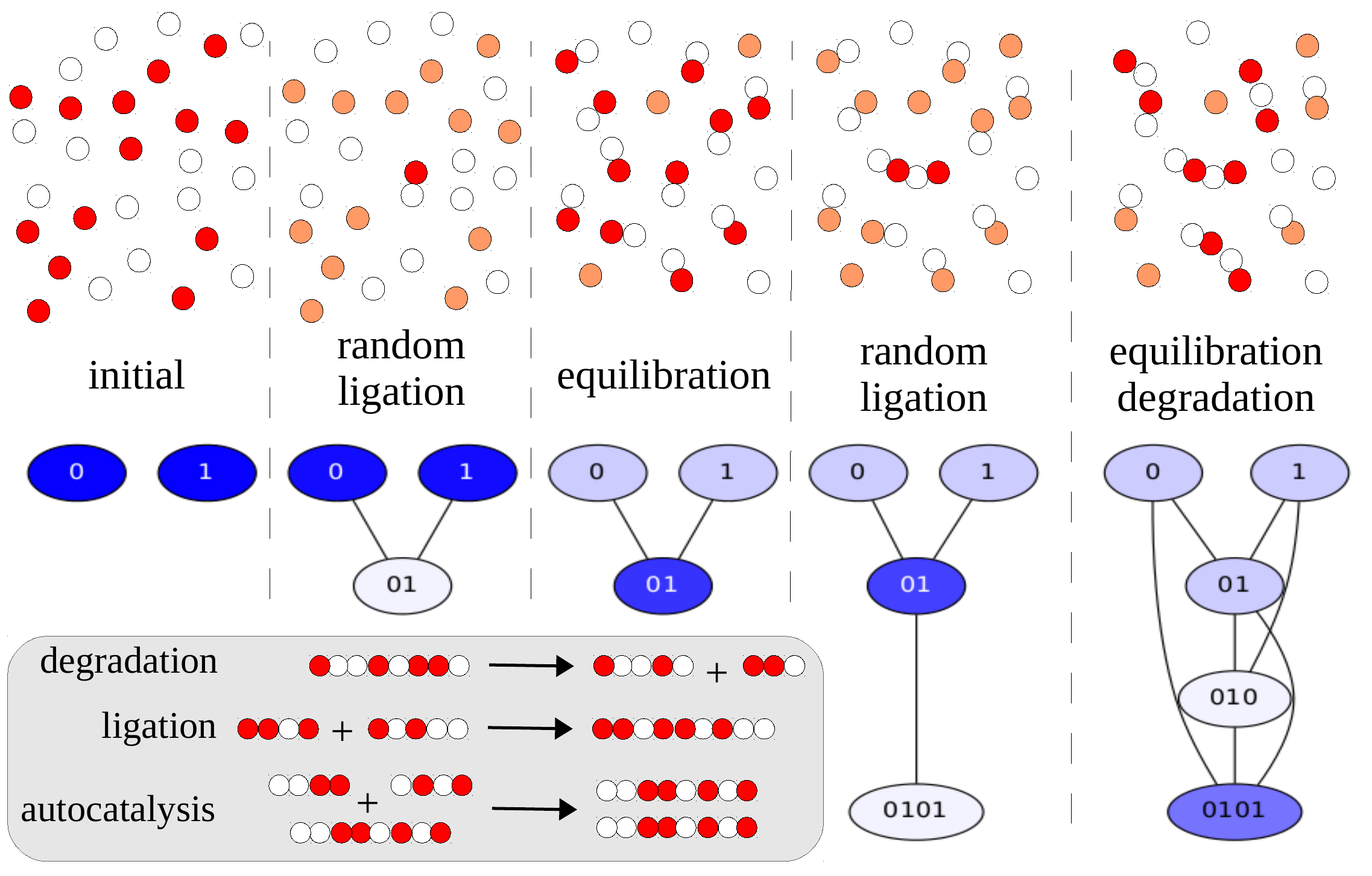}
\caption{
	In the \emph{exact autocatalytic polymer model} polymers made up of two
	types of monomers can degrade, ligate randomly or undergo autocatalytic
	replication (inlay).
	Strong autocatalysis and comparably rare random ligation promote the
	emergence of characteristic population structures (main image).
}
\label{fig_example}
\end{center}
\end{figure}

Several simple models have been developed to elucidate the prebiotic
emergence and evolution of nucleic acid replicators.
Most notably, random catalytic network models have been introduced to predict
the emergence of autocatalytic cycles in pools of arbitrary molecules with
random cross-catalytic
activity~\cite{stadler_random_1993,hanel_phase_2005,filisetti_stochastic_2012}.
More closely related to nucleic acids are catalytic polymer models, where
the catalyzed polymer ligation reaction is modeled as concatenation of random
strings~\cite{rasmussen_aspects_1985,farmer_autocatalytic_1986,
kauffman_autocatalytic_1986,rasmussen_emergence_1989,langton_spontaneous_1991,
fernando_stochastic_2007,hordijk_predicting_2012,derr_prebiotically_2012,
hordijk_autocatalytic_2016}.
Most studies of polymer models concern themselves with the appearance of
catalytic cycles when the catalytic activity of polymers is assigned either
randomly or dependent on the sequence around the ligation site.
Common conclusion of these studies is that random catalytic activity among a
set of molecules is sufficient to promote the appearance of reflexively
autocatalytic sets, i.e. groups of molecules that are able to jointly replicate 
themselves through cross-catalysis, see e.g. Refs.~\cite{kauffman_autocatalytic_1986,
hanel_phase_2005,hordijk_conditions_2014}.

In this study, we take the appearance of catalysis and replication for granted, 
and raise the critical questions:
do catalysis, replication, competition, cooperation, and selection indeed allow 
for the free exploration of the chemical sequence space or are some prebiotic 
evolutionary fates more common than others?
If the latter, what is driving the evolution in such chemical networks?
Ultimately, we want to better understand how prebiotic chemistry with reversible
reactions can give rise to the emergence of history, \emph{i.e.} can lead to a
series of path dependent events, where the outcome of a past selection events
impacts on what becomes accessible to evolution in the future.

In order to focus on these questions, we take a radical stance and study a
polymer model in which autocatalysis is assumed to work perfectly for all
molecular species.
There are several reasons to do so.
Firstly, if we believe chemical evolution to have evolved more and more
effective self-replicators, it should be just as informative to understand the
limiting case of perfect autocatalysis as it is to understand the limiting
case of completely random chemistry.
Secondly, by demanding perfect autocatalysis, we are able to derive closed
analytic expressions for all involved molecular species.
These allow us to scrutiny the reaction system with unprecedented analytic
rigor, and we are able to align our simulation results with the analytic
derivations.

We have previously reported on this model of self-replicating binary polymers
where exact autocatalytic replication introduces a selection bias toward few
specific replicator motifs and population structures~\cite{tanaka_structure_2014}.
Here, we present a full analysis of this system and several prebiotically
motivated model variants.

In the remainder of the article, we give a detailed exposition of our simple
binary polymer model with exact autocatalysis.
We derive analytical results for the two important limiting cases of purely
random as well as purely autocatalytic replication.
We proceed by presenting simulation results of the prebiotically interesting
case, where autocatalysis is dominent but augmented with relatively rare random
ligation.
For this scenario we reveal a strong inherent selection pressure toward
repetitive sequence motifs.
We suggest two non-exclusive explanations for the strong observed selection
pressure, one that dominates the transient dynamics of the system, and one
that acts upon equilibrated populations.
We also analyze several variants of our model, including larger alphabet
sizes, mutations, sequence complementarity, and non-replicating food species,
before discussing the general implications of our findings.

\section{An exact autocatalytic binary polymer model}
We study a system of self-replicating polymers made up of two types of monomers.
Each polymer can (1) decompose into two substrands by hydrolysis of any of its
bonds, (2) randomly ligate with another polymer to form a longer polymer, and
(3) replicate itself autocatalytically by ligating two matching substrands.
As an idealization, we assume that the rate constants of these reactions are
constant and in particular independent of the sequence information of each
polymer.
As we are interested in the effect of catalysis, we assume that random ligations
are comparatively rare.

The resulting system is a highly coupled molecular ecology, where replicating
species compete for common resources and serve as each other's reactants and
degradation products.
What will typically happen in such an ecology when starting from a pool of
monomers?

To formalize this reaction system we first introduce some notation:
let $\mathcal A=\left\{0,1\right\}$ be a binary alphabet.
We denote the set of strings over $\mathcal A$ by $\mathcal A^*$, and we write
$|k|$ for the length of string $k \in \mathcal A^*$.
For $l, m \in \mathcal A^*$ we denote by $l.m$ the string obtained by
concatenating $l$ and $m$.

We can now define the above reaction system formally, using the following
three reaction rules:
\begin{align}
	\label{eq_decomposition}
	l.m  &\xrightarrow{c_0} l+m \\
	\label{eq_uncatalyzed}
	l+m &\xrightarrow{c_1} l.m \\
	\label{eq_catalyzed}
	l+m+l.m &\xrightarrow{c_2} 2 \; l.m .
\end{align}

It has been observed experimentally, that non-enzymatic replicators suffer from
product inhibition~\cite{von_kiedrowski_parabolic_1991}, where most potential templates are in
their inactive double strand configuration, and thus cannot serve as replication
templates.
As a consequence, experimental replicators do not follow the exponential growth
dynamics of simple autocatalysis and, in turn, promote coexistence of replicator
species, rather than survival of only the
fittest~\cite{szathmary_sub-exponential_1989,fellermann_growth_2011}.
In our study, the implicit energy flow entering reaction~\eqref{eq_catalyzed}
is assumed to separate the inactive and low energy double strand configuration
of replicator strands and to transform them into their activated single strand
configuration.

Since $\mathcal A^*$ is enumerable, we can take it as basis for an infinite
dimensional state space:
each state vector $x$ denotes a population within which every species
$k\in \mathcal A^*$ spans one dimension and its associated coordinate
$x_k \in \mathbb{R^+}$ specifies the concentration of the respective species.
Standard dynamical systems analysis is complicated by the infinite
dimensionality of this state space.
Previous studies circumvented this difficulty by introducing a maximal strand
length~\cite{langton_spontaneous_1991,hordijk_predicting_2012} or focusing on very small monomer
pools~\cite{wu_origin_2009,derr_prebiotically_2012}.
Here, we present a mathematical formalism that is able to deal with the infinite
dimensionality for the most part analytically.

Assuming mass action kinetics, we can derive dimensionless reaction kinetic
equations for the molar concentrations $x_k$ of species $k \in \mathcal A^∗$ as
\begin{align}
	\frac{\mathsf dx_k}{\mathsf{dt}} &= \phantom{\alpha}\left(\sum_{\substack{k.j=i \\ j.k=i}}x_i-\sum_{i.j=k}x_k\right) \nonumber \\
	&+ \alpha\left(\sum_{i.j=k}x_ix_j-\sum_{\substack{k.j=i \\ j.k=i}}x_jx_k\right) \nonumber \\
	&+ \beta\left(\sum_{i.j=k}x_ix_jx_k-\sum_{\substack{k.j=i \\ j.k=i}}x_ix_jx_k\right)\equiv f_k.
	\label{eq_kinetic}
\end{align}
where we have chosen a time scale $t = c_0^{-1}$, and introduced the
non-dimensional random ligation rate $\alpha=c_1/c_0$, and the non-dimensional
autocatalytic ligation rate $\beta=c_2/c_0$.
The above summations consider every pair of species $i$ and $j$ which
satisfy $i.j=k$, and $k.j=i$ or $j.k=i$.
Formally, the sum operators are defined by means of the indicator functions
($\mathbf 1_{i.j=k} = 1 \text{ if } i.j=k \text{ else } 0$) as follows:
\begin{align}
	\sum_{i.j=k} &:= \sum_{i,j \in \mathcal A^*} \mathbf 1_{i.j=k} \\
	\sum_{\substack{j.k=i \\ k.j=i}} &:= \sum_{i,j \in \mathcal A^*} (\mathbf 1_{j.k=i}+\mathbf 1_{k.j=i})
\end{align}
Note that the second definition may involve an infinite sum and is only defined
if the right-hand-side is defined.
For finite size systems, which we consider in this article, the sum is always
defined and allows us to write down the reaction equation system in a compact
closed form without introducing any arbitrary truncation.

Equation~\eqref{eq_kinetic} allows us to perform standard stationarity and
stability analysis for the two limiting cases of purely random ligation
as well as purely autocatalytic ligation.

\subsection{Purely random and purely autocatalytic limit cases}
\label{sec_model}

In the absense of autocatalysis ($\beta=0$), the system approaches a stationary
and stable equilibrium distribution, where the concentration of each replicator
decays exponentially with its length:
\begin{equation}
	\left(x_\text{exp}^*\right)_k = \frac 1 \alpha \me^{-b|k|},
	\label{exponential}
\end{equation}
which can be readily confirmed by inserting~\eqref{exponential}
into~\eqref{eq_kinetic} for $\beta=0$.
Here, $b$ is a constant determined by the boundary condition.
Importantly, there is no selection for specific sequence patterns.

We will now show that this situation changes radically in the limit of purely
autocatalytic ligation ($\alpha=0$).
Firstly, since decomposition is now the only reaction that introduces novel
species, any stationary state can only consist of initially present strands or
their substrands.
For any strand that is populated in the stationary state, all its substrands
must be populated as well.
This can be proven by contradiction: assume there exists a steady state $x^*$
with $x_m^*>0$ but $x_k^*=0$ for some $k.j=m$.
From equation~\eqref{eq_kinetic} it follows:
\begin{equation}
	\left.\frac{\mathsf dx_k}{\mathsf{dt}}\right|_{x^*} = \sum_{\substack{k.j=i \\ j.k=i}} x_i^* \geq x_m^* > 0.
\end{equation}
Since $f_k(x^*) > 0$, $x^*$ cannot have been a steady state.
Thus, the assumption is false and species $k$ must be populated as well.

We call any strand that is not a substrand of any other strand in the population
a ``chief'' and the set of its substrands its ``clan''.
An example of such a ``chief-clan'' structure is shown in
figure~\ref{fig_example}.
We denote the set of all chiefs of $x$ by $\partial\mathcal{A}^\dagger_x$ and
the set of all clan members by $\mathcal{A}^\dagger_x$.
Using this nomenclature, the stationary chiefs of the purely autocatalytic case
are those substrands of initally present strands that survived decomposition
and are not substrands of other surviving species.

Interestingly, under our assumption of exact replication~\eqref{eq_catalyzed},
stationary species concentrations no longer decay exponentially with their
length.
Instead, any proper clan member is maintained at a constant concentration given
by the replication rate constant $\beta$:
\begin{align}
	x^*_k&=
	\begin{cases}
	\beta^{-1/2}, & k\in \mathcal{A}^\dagger_x \backslash \partial\mathcal{A}^\dagger_x  \\
	0, & k \not\in \mathcal{A}^\dagger_x.
	\end{cases}
\label{constant}
\end{align}
This can again be readily confirmed by inserting~\eqref{constant}
into~\eqref{eq_kinetic} with $\alpha=0$.

Concentrations of chief species $m \in \partial\mathcal A^\dagger_x$ are
unconstrained in the steady state, appart from what is determined by the
boundary condition.
This is so because the steady state condition for chiefs
\begin{multline}
	\left.\frac{\mathsf dx_m}{\mathsf{dt}}\right|_{x^*}
	= - \sum_{i.j=m} x_m^* + \beta\sum_{i.j=m} x_i^*x_j^*x_m^* \\
	= x_m^* \left(\beta\sum_{i.j=m} \beta^{-1} - \sum_{i.j=m} 1\right) = 0
\label{eq_chief}
\end{multline}
is always fulfilled, independent of $x_m$.

\label{sec_linear_stability}
To understand how stationary solutions respond to fluctuations, we derive the
entries of the functional matrix:
\begin{multline}
	\left.\frac{\partial f_k}{\partial\mathsf x_l}\right|_{x^*} = \sum_{\substack{k.j=i \\ j.k=i}}\frac{\partial x_i}{\partial \mathsf x_l} - \sum_{i.j=k} \frac{\partial x_k}{\partial \mathsf x_l} \\
	\quad + \beta \left(\sum_{i.j=k}\frac{\partial (x_ix_jx_k)}{\partial \mathsf x_l} - \sum_{\substack{k.j=i \\ j.k=i}} \frac{\partial (x_ix_jx_k)}{\partial \mathsf x_l}\right) \\
	= \left(\sum_{\substack{j.l=k \\ l.j=k}} 1 - \sum_{\substack{k.l=i \\ l.k=i}} 1 - \delta_{kl}\sum_{\substack{k.j=i \\ j.k=i}} 1 \right) ,
\end{multline}
where $\delta_{ij} = 1$ if $i=j$ else 0 is the Kronecker symbol,
and chief concentrations are assumed to be $\beta^{-1/2}$ for simplicity.

The equation simply expresses a detailed balance relation:
increasing the amount of any clan species results in a flux to any species that
uses the increased strand as prefix or suffix, while simultaneously introducing
a flux away from all species needed as reactants for the formation of those
longer species.
Again, increasing the level of chief species ($l=m$) does not introduce any
flux in the population structure.
Notably, the stability of a stationary population is entirely determined by the
topology of the reaction graph, not by the reaction rates.

To fully determine the stability of stationary solutions requires to solve for
the eigenvalue spectrum of the functional matrix.
The analytical treatment is complicated by the fact that the dimension of our
system is formally infinite.
We therefore resolve to numerically probing the eigenvalue spectrum of hundreds
of different trial populations (see table~\ref{table_chiefs}).
We find that the vast majority of stationary chief-clan structures have indeed
no positive eigenvalues.
Moreover, we find that the number of zero eigenvalues either equals the
number of chiefs in the population or is one greater than that.
Up to two of these zero eigenvalues result from mass conservation.
The remaining ones express the fact that chief-clan structures with competing
chiefs are connected by lines of linearly stable solutions.
For example, in a population with competing chiefs $01$ and $10$ all populations
that satisfy the constant solution $x_0=x_1=x^*$ are linearly stable,
independent of how material is distributed among the chiefs.

\begin{table}[t]
 \begin{center}
  \caption{Stability of clans with different number of chiefs.}
  \label{table_chiefs}
  \begin{tabular}{p{1.0cm}p{1.5cm}p{1.5cm}p{1.0cm}}
  \multicolumn{4}{c}{} \\
  \multicolumn{4}{c}{Clans with one chief} \\\hline
   $|m|$ & $P_{\rm sn}$ & $P_{\rm ss}$ & $P_{\rm u}$ \\
   \hline
   2  & 1.00 & 0.00 & 0.00 \\
   4  & 1.00 & 0.00 & 0.00 \\
   6  & 0.40 & 0.60 & 0.00 \\
   8  & 0.20 & 0.80 & 0.00 \\
   10 & 0.07 & 0.93 & 0.00 \\
   12 & 0.02 & 0.98 & 0.00 \\
   14 & 0.01 & 0.99 & 0.00 \\
   16 & 0.00 & 1.00 & 0.00 \\
   18 & 0.00 & 1.00 & 0.00 \\[2mm]
   \hline
  \multicolumn{4}{c}{} \\
  \multicolumn{4}{c}{Clans with two chiefs} \\\hline
   $|m|$ & $P_{\rm sn}$ & $P_{\rm ss}$ & $P_{\rm u}$ \\
   \hline
  2  & 1.00  & 0.00 & 0.00 \\
  4  & 0.72  & 0.28 & 0.00 \\
  6  & 0.21  & 0.79 & 0.00 \\
  8  & 0.04  & 0.95 & 0.00 \\
  10 & 0.00  & 0.98 & 0.02 \\
  12 & 0.00  & 0.97 & 0.03 \\
  14 & 0.00  & 0.95 & 0.05 \\
   \hline 
  \multicolumn{4}{c}{} \\
  \multicolumn{4}{c}{Clans with four chiefs} \\\hline
   $|m|$ & $P_{\rm sn}$ & $P_{\rm ss}$ & $P_{\rm u}$ \\
   \hline
     2  & 1.00 & 0.00 & 0.00 \\
     4  & 0.82 & 0.18 & 0.00 \\
     6  & 0.19 & 0.81 & 0.00 \\
     8  & 0.01 & 0.99 & 0.00 \\
     10 & 0.00 & 0.99 & 0.01 \\
     12 & 0.00 & 0.99 & 0.01 \\
   \hline
  \end{tabular}\\   
  \begin{tabular}{p{8cm}}\footnotesize
	$|m|$: the length of the chief(s), $P_{\rm sn}$: the probability to find
	stable nodes, $P_{\rm ss}$: the probability to find stable spirals,
	$P_{\rm u}$: the probability to find unstable stationary points.
	The number of sampling was $10^4$.
	Only combinations of chiefs that had the same number of ``0'' and ``1'' were
	sampled.
  \end{tabular}
 \end{center}
\end{table}

We remind that linear stability analysis only detects local stability and does
not discriminate the relative stability of one solution over another.
Moreover, our analysis does not consider boundary conditions, which might render
some chief-clan structures less likely than others, if not outright impossible.

How many potential equilibrium solutions are there?
We can roughly estimate the potential stable chief-clan structures as a function
of the longest population member as follows:
there are $2^{|m|}$ different chiefs of length $|m|$.
If any of these chiefs can be either absent or present in the population, then
there are $2^{2^{|m|}}$ different potential steady state populations.
Despite the constraints of equation~\eqref{constant}, there are
super-exponentially many stable potential equilibrium populations.
Already for strands as short as ten, this amounts to more than $10^{300}$
possible stationary chief-clan structures.
One should bare in mind, however, that this estimation ignores the boundary
condition imposed by the initial (and conserved) material amount.

So far our analysis has been based on molar concentrations.
If material is sparse, however, it is more accurate to adopt the view that
molecular species are present in discrete amounts.
Under this view, the continuous equations
\eqref{eq_decomposition}--\eqref{eq_catalyzed} still describe the expected
tendencies of the dynamic process, but with the difference that species can
become extinct and disappear entirely, rather than being diluted to arbitrary
small (but non-vanishing) concentrations.
Under pure autocatalysis, this is especially relevant for chief species, as
they are not constrained to potentially high equilibrium amounts.
If chief species disappear as the result of random fluctuations, they become
permanently extinct.

\subsection{Combined random and autocatalytic ligation}
We now study systems where both ligation reactions act in combination.
In particular, we analyze \emph{how} the system transitions from the
exponential equilibrium solution to stable chief-clan structures when varying
the ratio of the random and autocatalytic ligation rates.

To answer this question, we sample the stochastic process defined through
\eqref{eq_decomposition}--\eqref{eq_catalyzed} via exact stochastic
simulation using a Gillespie algorithm.
In order to tackle the dimensionality of the problem, the algorithm
generates for the current state $x(t)$ a list of potential follow-up states
on the fly and evaluates transition rates $r_i$ according to
equation~\eqref{eq_kinetic}.
For any finite size population, this transition list will also be finite.

Simulations are started from a monomer pool with
$x_0(0)=x_1(0)=2100$, $\beta=10^{-4}$ and $\alpha$ varying from $10^{-7}$
to $10^{-3}$.
Under sufficiently rare random ligation, we expect these settings to stabilize
chief-clan structures up to length ten, and under only random ligation, an
exponential distribution up to length eight.
With the given $\beta$, in a population distributed according
to~\eqref{constant}, random and autocatalytic ligation would be equally
likely for a random ligation rate constant $\alpha=\beta^{-1/2}=10^{-2}$.

After $t=1000$ time units, populations are compared to both equilibrium
solutions~\eqref{exponential} and~\eqref{constant} using the cosine similarities
\begin{equation}
	s_\text{exp} = \frac 1 \alpha\frac{\sum_{k\in\mathcal A^*} x_k \mathrm e^{-b|k|}}{\left|x\right|\left|x_\text{exp}^*\right|} 
\end{equation}
and
\begin{equation}
	s_\text{ccs} = \frac{\sum_{k \in \bar{\mathcal A_x}} x_k \beta^{-1/2}}{|\sum_{k \in \bar{\mathcal A_x}} x_k|\left|\sum_{k \in \bar{\mathcal A_x}} \beta^{-1/2}\right|}
\end{equation}
where $\bar{\mathcal A_x} = \mathcal A^\dagger_x \backslash \partial \mathcal A^\dagger_x$.
$s_\text{exp}$ the distance toward the corresponding exponential equilibrium and
$s_\text{ccs}$ measures the distance to the smallest enclosing chief-clan
structure, \emph{i.e.} the smallest chief-clan structure that has a non-zero
species count for any species that is present in the population.
\begin{figure}
	\includegraphics[width=\columnwidth]{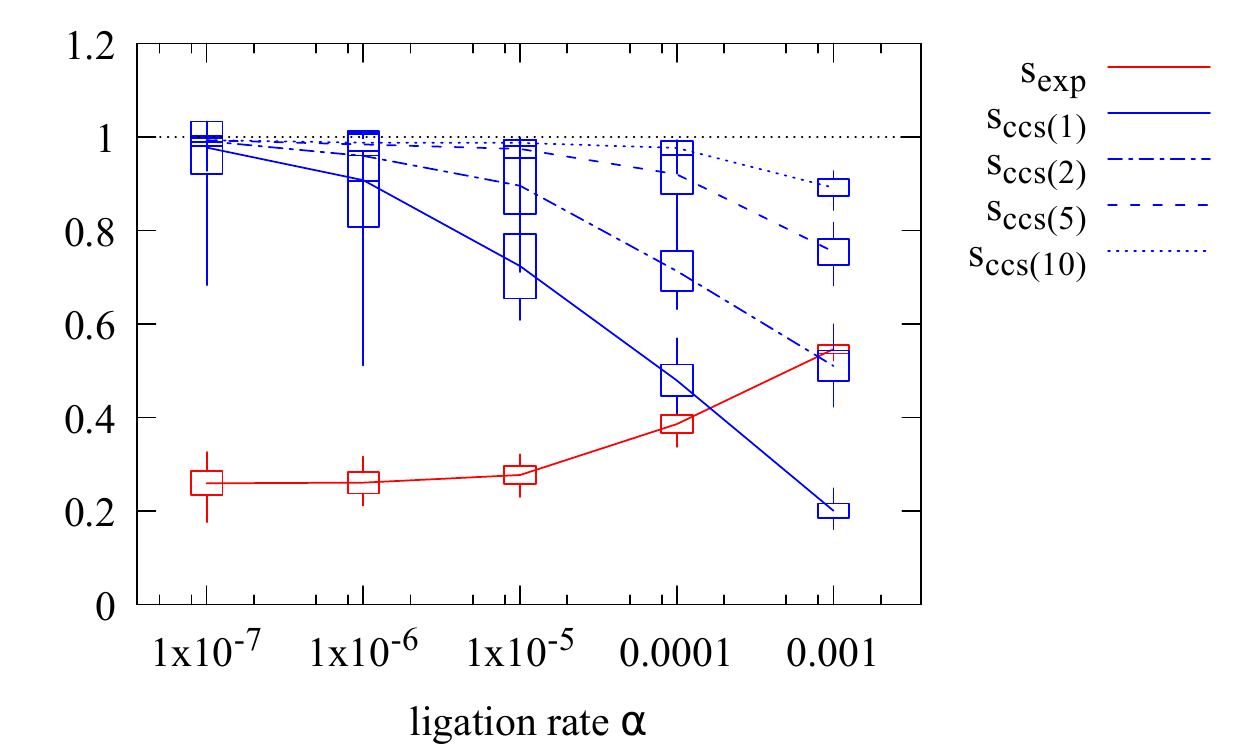}
	\caption{
	Cosine similarity between simulated population samples and ideal exponential
	distributions (red) and between the samples and their smallest enclosing
	chief-clan structure (blue), the latter for different threshold values.
	Species with occupancies below the threshold are not considered for the
	distance calculation.
	}
	\label{fig_transition}
\end{figure}
Fig.~\ref{fig_transition} shows the transition from stable chief-clan
structures into random distributions.
With increasing random ligation rate $\alpha$, observed populations seize to
resemble their original chief-clan structure.
Visual inspection of the results revealed that these deviations are mainly
caused by few random ligation products that are not numerous enough to establish
their clan.
When filtering ouf such outliers, we found that the remaining core populations
resemble more closely their corresponding chief-clan structure, essentially
until $\alpha$ reaches $10^{-2}$.
At the same time, populations approach equilibrium distributions, because
autocatalysis becomes too slow to generate chief-clan structures for the novel
species created by random ligation.
For high enough $\alpha$, deviation from the exponential equilibrium distribution
is again mainly caused by rare long strands, whose expectation number from
equation~\eqref{exponential} is well below a single molecule.

These results indicate that chief-clan structures cannot develop above
a critical random ligation rate less than $\beta^{-1/2}$.
Approaching this critical threshold, chief-can structures become
succeedingly obscured by random ligation products in low copy number.

\subsection{Rare random ligation under strong autocatalysis}

With the above insights we now concern ourselves with the prebiotically
interesting parameter regime where comparatively rare random ligations
occasionally introduce novel species into populations with strong autocatalytic
replication.
When starting from a pool of monomers, random ligation will create longer
species that get amplified via autocatalytic replication and generate their
clans via decomposition.
We expect this process to generate chief-clan structures satisfying equation
\eqref{constant} of a size that is just able to accommodate the overall amount
of material.

Our question now becomes:
does autocatalytic replication with rare random ligation generate all of the
overwhelmingly many potential chief-clan structures, or is there some
reason for some chief-clan structure to be more prominent than others?

\begin{figure}
\begin{center}
\includegraphics[width=\columnwidth]{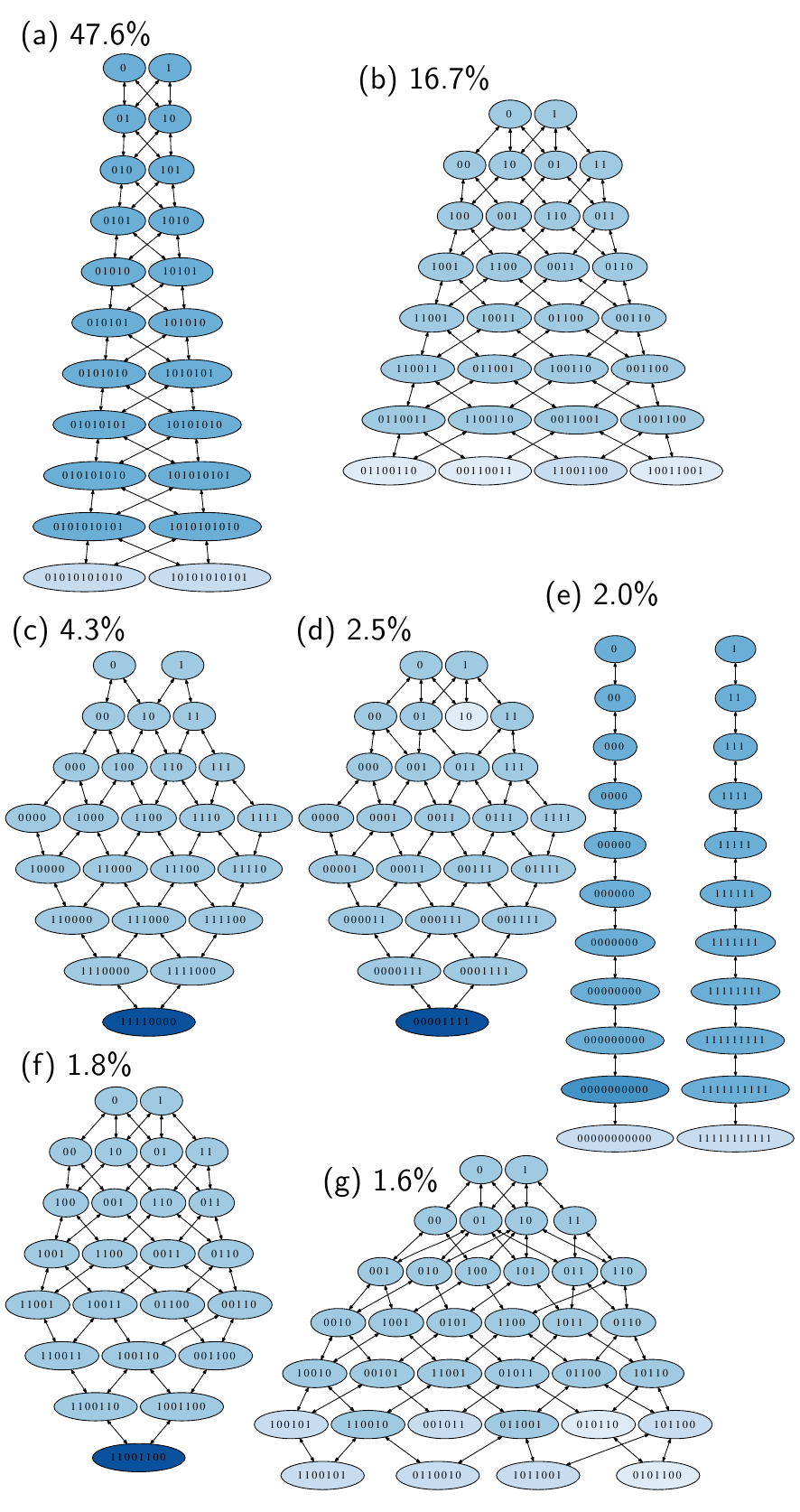}
\caption{
	The result of hierarchical clustering for 1000 populations
	produced by simulation; Named: (a) bootlace, (b) pinecone1, (c)
	pinecone2, (d) pinecone2, (e) two-towers, (f) pinecone1.
	Cluster (g) has no name and there are many similar structures.
	Note that (b) and (f), and, (c) and (d) belong to the same symmetry.
	These seven clusters occupy 76.5\% of all the population produced.
	Species having less than 10\% of the constant solution are omitted.
	Simulations were run with $\alpha=10^{-10}$, $\beta=10^{-7}$, and
	$x_0(0)=x_1(0)=200\,000$.
}
\label{fig_selection}
\end{center}
\end{figure}

Using the results of 1000 simulation runs (simulated until $t=100$), we
perform single linkage hierarchical clustering based on cosine distance
\begin{equation}
	d(x,x') = 1 - \frac{\sum_{k\in\mathcal A^*} x_kx_k'}{|x||x'|}.
\end{equation}
We find that about 75\% of
the simulation results concentrate in six clusters with characteristic
population structures, shown in figure~\ref{fig_selection}.
This is a very strong selection pressure, given that all our kinetic equations
are sequence agnostic---especially under the light of the results of the
previous discussion.

Interestingly, the selected populations have very few chiefs composed of short
repeating sequence motifs such as 01 in cluster \ref{fig_selection}(a)
(bootlace), 0011 in cluster \ref{fig_selection}(b) and (f) (pinecone1), 00001111
and 11110000 in cluster \ref{fig_selection}(c) and (d) (pinecone2).
In about 2\% of the simulation runs, we even observe complete demixing of the
monomers into homopolymers composed exclusively of 0's and 1's respectively
(cluster \ref{fig_selection}(e), twotowers).

We also observe that some clusters exhibit an ``inverted'' population structure
where chiefs are more populated than other clan members
(\ref{fig_selection}(c), (d), and (f)).
This is due to the boundary condition where the total material amount is too
low to maintain a longer chief but too high for the chief concentration to be
comparable to the equilibrium value.
As a consequence, the highly populated chief will more likely create---via
random ligation---species outside its clan structure, which either die out or
cause the whole population to transition into another structure.
Indeed, we have observed in simulation that these metastable structures exhibit
stronger random fluctuations and potentially transition into completely distinct
chief-clan structures~\cite{tanaka_structure_2014}.

We emphasize that the stochastic simulation results are rather unexpected, if
not even in apparent contradiction to the results of the linear stability
analysis derived above. 
If linear stability theory allows for almost any chief-clan structure to be (at
least meta-)stable, how come that stochastic simulation exhibits such a strong
bias toward selecting very few and very regular population structures?
What constitutes this ``survival of the dullest''?

\section{Mechanisms of structure selection}
\label{sec_structure_selection}
We now propose two possible explanations for the symmetry breakdown in this
replicator ecology:
On the one hand, we identify a selection bias in the \emph{transient} process of creating
equilibrated, non-inverted chief-clan structures.
On the other hand, we identify a selection bias in the \emph{steady state} dynamics
of an established equilibrated population.
Both selection biases steer the system toward highly repetitive sequence
patterns, and we hypothesize that both of them are responsible for the strong
selection observed in stochastic simulation.

\subsection{Selection in locally equilibrated populations}
We first discuss the equilibrium scenario.
Consider an equilibrated population structure that satisfies~\eqref{constant}
and comprises two chiefs $a$ and $b$.
The clans of these chiefs always overlap (because monomers take part in all
clans) and the overlap forms a pool of common resources for the two clan
structures to compete over.
As in the end of the previous section, we assume random ligations to be
sufficiently rare and use equilibrated chief-clan structures as initial
conditions.

It is illustrative to first investigate a minimal system of four species, where
two chiefs $01$ and $10$ compete over the shared resources $0$ and $1$.
We know from the above that mass conservation of the two monomers fixes this
system on a two dimensional plane, which features a continuous line of
marginally stable steady states due to condition~\eqref{constant}.
The line connects the two boundary states, formed by populations with only one
surviving chief species.


In a stochastic treatment of this system, the line of connected equilibria,
translates into a drift term that vanishes for equilibrium solutions.
Unlike in the deterministic treatment, random fluctuations can now redistribute
material along the line of marginal steady states with virtually no deviation
from equilibrium.
The two boundary states are absorbing boundaries, as the walk terminates if one
of the competing chiefs becomes extinct (and cannot be repopulated in
the absense of random ligations).

Interestingly, diffusion of the random process also becomes position dependent:
randomly converting a single chief molecule from one species to the other
requires the abundance of both chief species, one entering as reactant to be
degraded, the other as catalyst to enhace its own ligation.
Diffusion is therefore the highest when the product of the two chiefs is
maximized and goes toward zero when approaching the boundaries, where either
of the two species becomes sparse.

In the appendix, we develop a one dimensional random walk, constrained to the
line of connected steady states, that approximates the two dimensional case.
For this approximation, we are able to analytically derive a Fokker-Planck
equation that features zero drift and position dependent diffusion:
\begin{align}
	\frac{\partial}{\partial\mathsf t} P(x_{01},t)
	&\approx \frac{\partial^2}{\partial\mathsf x_{01}^2} \left[D(x_{01}) P(x_{01},t)\right]
	\label{eq_FP}
\end{align}
where $D(x)=\frac{x_{01}x_{10}-1}{x_{01}+x_{10}}$.

The parabolic diffusion term has a remarkable effect on the first hitting time,
\emph{i.e.} the typical time of chief coexistence.
Because diffusion is higher in the center of the manifold (where both chiefs
are populated about evenly) the system is more prone to leave this state
and move toward states of lower diffusion.
This diffusion induced effective drift drives the system toward regions where
one chief outnumbers the other.
The average time for the system to reach one of the absorbing states is
therefore orders of magnitude shorter than for a random walk process with
comparable, but constant diffusion.
Figure~\ref{fig_hitting_times} of the appendix compares the cumulative probabilty
functions of the hitting time distributions for the original two dimensional
system, our one dimensional approximation, and a random walk process with
constant diffsuion.

Perheaps contrary to expectations raised  by the previous linear stability
results, we now see that there is a natural tendency for minority species to be
driven into extinction.
This effective drift is ultimately induced by position dependent diffusion
along the line of competing chief populations.

Note that this effective drift can only be observed among chiefs that can
absorb all the material of their competitor:
otherwise, both chiefs can survive.
For example, two chiefs $010$ and $101$ can coexist because neither can absorb
all the material of the other.
Yet, deviations in the chief-clan structure from the constant solution can
sometimes absorb a chief with different material than its competitor, as we
discuss now.

\begin{figure}
\begin{center}
	\raisebox{4.2cm}{(a)}\includegraphics[width=5cm]{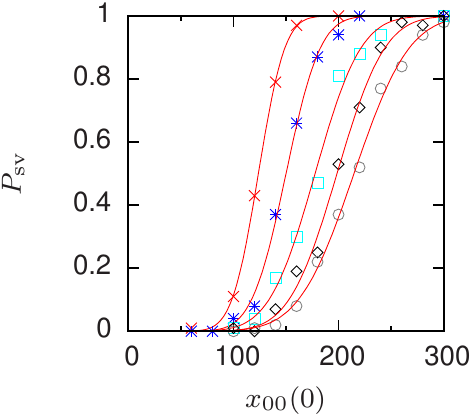}\\[2mm]
	\raisebox{4.2cm}{(b)}\includegraphics[width=5cm]{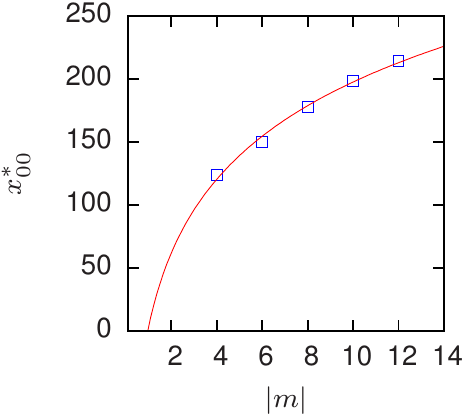}\\
\caption{
	(a) Probability of survival, $P_{\rm sv}$, after $t=1000$ for 00 coexisting
	with a bootlace structure (figure~\ref{fig_example}) with different chief
	length, depending on its initial concentration $x_{00}(0)$.
	The chief length are: $|m|=4$ ($\times$), $|m|=6$ ($\ast$), $|m|=8$
	($\square$), $|m|=10$ ($\diamond$), and $|m|=12$ ($\circ$).
	The solid curves are fitted error functions.
	(b) The threshold value, $x_{00}^\ast$, where $P_{sv}=0.5$ against the chief
	length $|m|$.
	The solid curve is a fitted logarithmic function.
}
\label{fig_competition2}
\end{center}
\end{figure}

Figure~\ref{fig_competition2} shows how a short chief, $00$, can be driven into
extinction as the result of competion with longer chiefs:
simulations are initialized with equilibrated bootlace structures of varying
length, plus $00$, and measure the probability of survival of the shorter chief
after 1000 time units as a function of its initial concentration for varying
length of the longer chief. In this case, chiefs in a bootlace structure,
such as $0101$, cannot absorb all the material of $00$, but the short chief
can nevertheless be absorbed by fluctuations of the chief-clan structure, which
in turn deviates from the exact solution~\ref{constant}.
The simulation shows the clear tendency of extinction of shorter clans in
competition.

While the above reveals the source of competition and extinction in our model,
and indicates that simulations should generate populations with few chiefs,
it does not explain why the dynamics select those particular chiefs that
feature simple repetitive motifs.

\begin{figure}[t]
	\includegraphics[width=\columnwidth]{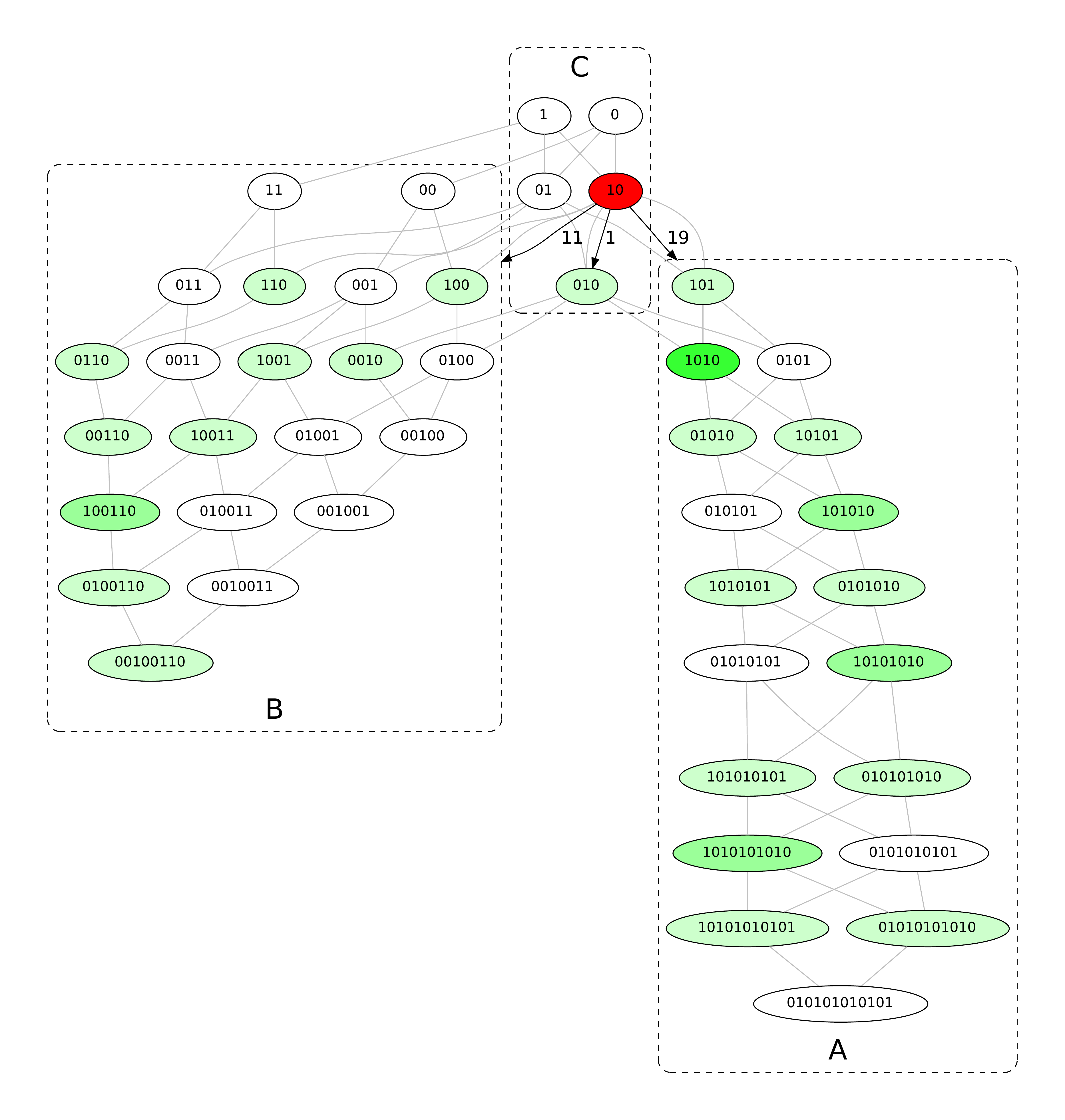}
	\caption{
	Visualization of the ``partial flows'' when a population in equilibrium
	state $x^*$ is perturbed by adding $\Delta x$ to species $l=10$.
	In the case shown, ${\md f_l}/{\md x_l} = -31$ (red).
	This flow is distributed such that $-{\md f^A_l}/{\md x_l} = 19$ go into
	clan $A$, $-{\md f^B_l}/{\md x_l} = 11$ go into clan $B$, and
	$-{\md f^C_l}/{\md x_l} = 1$ remain in the overlap $C$.
	Green nodes show the recipients of material where light green corresponds
	to $1$, medium green to $2$, and dark green to $4$.
	}
	\label{fig_competition}
\end{figure}

An answer to this question lies in the way perturbations are distributed
among competing clans, as shown schematically in figure~\ref{fig_competition}:
When species in the overlap of competing clans are perturbed through random
decomposition of longer strands, detailed balance causes an opposing flow from
the perturbed species toward longer strands.
This outflow can be partitioned into three components:
material flowing into species belonging exclusively  to clan $A$, clan $B$, or
their overlap $C$:

\begin{equation}
	\frac{\partial f^{A \cup B \cup C}_l}{\partial \mathsf x_l} = \frac{\partial f_l}{\partial \mathsf x_l}.
\end{equation}
where we have introduced the ``partial'' flow
$\frac{\partial f^A_l}{\partial \mathsf x_l}$ from $l$ into the species in $A$:
\begin{equation}
	\left.\frac{\partial f^A_l}{\partial \mathsf x_l}\right|_{x^*} = - \left(\sum_{\substack{l.j\in A \\ j.l \in A}} 1 + \sum_{l.l\in A} 2 \right) \approx -\sum_{\substack{l.j\in A \\ j.l \in A}} 1.
\label{eq_competition}
\end{equation}

Thus, the flow into $A$ ($B$) scales with the number of strands in $A$ ($B$)
that have the perturbed strand as their prefix or suffix.
Since regular chiefs can form longer strands with more repetitive sequences than
irregular chiefs, we expect regular chiefs to outperform irregular ones in
this competition.

\paragraph*{Summary}
In equilibrium strands with high population numbers outcompete low
populated strings in competition for resources and this makes it
difficult for randomly created chiefs to survive. In some cases
this process can be described by a Fokker-Planck equation, which
expresses a position dependent diffusion. Further, long clans can absorb
material faster than short clans because they offer more reaction
pathways to absorb fluctuations (due to an increased number of
prefix/suffix matches). With a given amount of material, 
regular chiefs can form longer clans than irregular ones.

\subsection{Selection far from equilibrium}

We now discuss the transient dynamics of a typical process of structure
generation that starts from a pool of monomers, where---as before---random
ligation is small enough for the system to locally equilibrate after each random
ligation event.
Thus, after spontaneous formation of an intermediate chief through random
ligation, the new chief will accumulate material via autocatalysis in order to
satisfy the equilibrium condition~\eqref{constant}.
The result is an inverted population structure where chief species accumulate
the majority of the material.
Therefore, the next ligation is more likely to happen among two chiefs than
among any other species in the population---thus creating a strand with
repeated sequence patterns~\cite{tanaka_structure_2014}.
Irregular chiefs, on the other hand, require additional substrands to be formed
from now less abundant clan members.
The process repeats until chief concentrations become lower than the equilibrium
solution, at which point there is no surplus material to grow longer strands.

The above verbal argument can be refined into a quantitative estimation of
selection likelihoods:
assuming that populations grow by irreversible random ligation and subsequent
equilibration, the likelihood to obtain one population from a previous one via
a single ligation event is proportional to the equilibrium abundance of the
reactants, which is largely determined by the equilibrium
condition~\eqref{constant}.
Assuming that the surplus material is evenly distributed among chief species,
we can calculate the probability to obtain any final population from the initial
condition by multiplying the conditional probabilities of ligations along each
ligation pathway that can lead to the final population in question.
However, because of the double exponential scaling of potential chief-clan
structures  as a function of chief length, the reaction system leads to a
combinatorial explosion that fastly exceeds computing resources.
Justified by equation~\eqref{eq_competition} and the results of
figure~\ref{fig_competition2}, we introduce the additional assumption that a
new chief can only form in a population if it is at least as long as any
existing chief.
With this additional assumption, the combinatorial problem can be tackled.

\begin{figure}[t]
\includegraphics[width=\columnwidth]{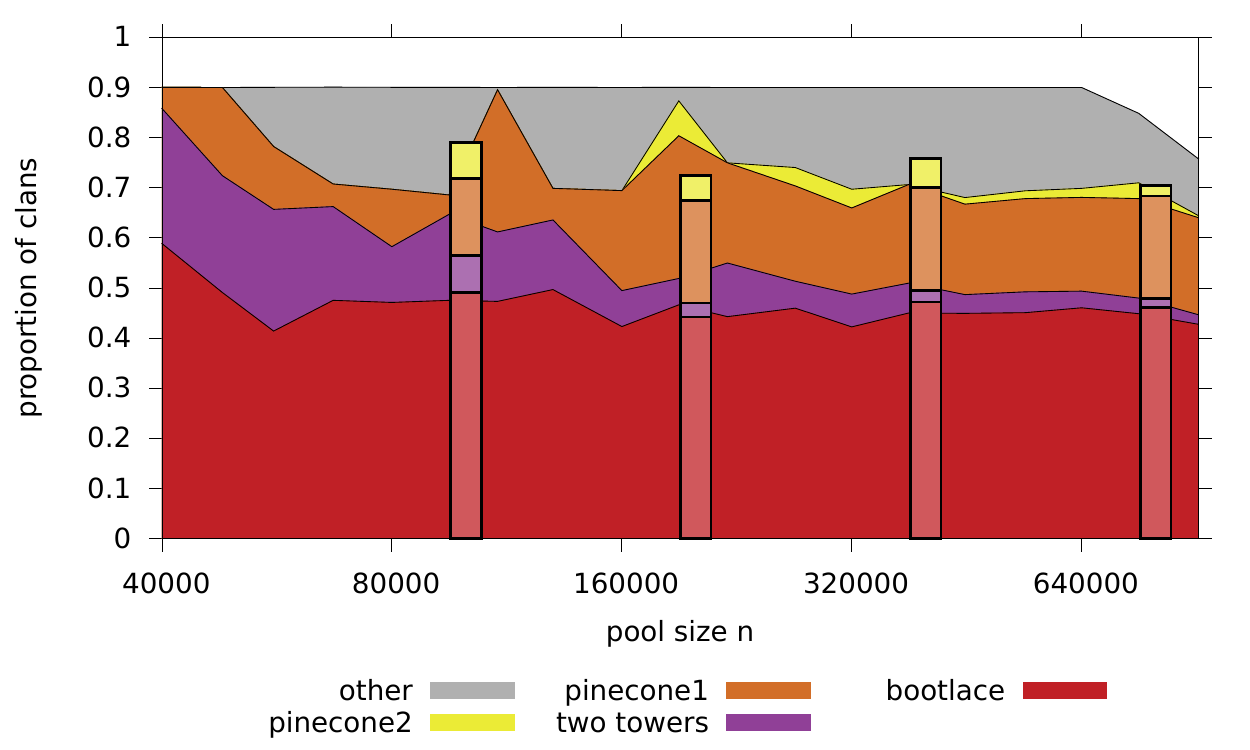}
\caption{
	Estimated likelihood to form populations of strands with certain motifs as
	a function of the monomer pool size with balanced monomer compositions.
	Note the exponential scale of the $x$ axis.
	Overlaid bar charts show results of full stochastic simulations for various
	system sizes.
	See text for details.
}
\label{fig_chief_likelihoods}
\end{figure}

We systematically perform these likelihood estimations for different initial
monomer amounts and estimate the 90\% most likely ligation pathways.
The resulting population structures are classified according to their sequence
motifs and compared to the results of stochastic simulations for selected pool
sizes.
The computed likelihoods---shown in
figure~\ref{fig_chief_likelihoods}---are generally in good agreement with the
simulations.

\paragraph*{Summary}
During the transient equilibration moves most material into chiefs.
If there is a single chief, it will most likely react with itself to create the
next chief (scenario $01+01\to 0101$).
The key here are inverted population structures~\cite{tanaka_structure_2014}
where most material is in the chief species.
To make irregular sequences during the transient, distinct building blocks are
needed.
If one chief has absorbed the material, there is less material for the other
chief, thus reducing the propensity of forming the irregular chief
(scenario $10 + 0 \to 100$).
Again most material is in the chief.  To make a particular irregular chief
during the transient, the two substrings have to ligate in correct order
(scenario $10 + 0 \to 100$ vs. $0 + 10 \to 010$), so the symmetry of the
reaction network plays as only one of two possible reactions produce a
non-repetitive sequence.
Finally, regular chiefs require the build-up of fewer intermediate species,
i.e. 01010101 requires only 01 and 0101, whereas 01110001 would require
01, 11, 00, 01, 0111 and 0001 or a similar number of other intermediates.

\subsection{Distinguishing the alternative selection mechanisms}
\label{sec_distinguishing}
We have presented two alternative driving mechanisms that might explain
the selection of few chiefs with short repetitive motifs.
Which of these mechanisms is responsible for the results presented in
figures~\ref{fig_transition} and~\ref{fig_selection}?

In order to discriminate the two selection mechanisms, we perform numerical
simulations and measure---over the course of time---the number of species
and chiefs, an activity measure $A$ that measures the surplus of material in
chief species through the definition
\begin{equation}
        A(x) = \sum_{x_i\in\partial \mathcal A^\dagger_x} \left(\frac{x_i}{x^*}\right)^2\left(\frac{x_i}{x^*}-1\right)^2 ,
\end{equation}
as well as the cosine similarity $s_\text{ccs}$ of the population with the
smallest enclosing chief-clan structure.

\begin{figure}
	\includegraphics[width=\columnwidth]{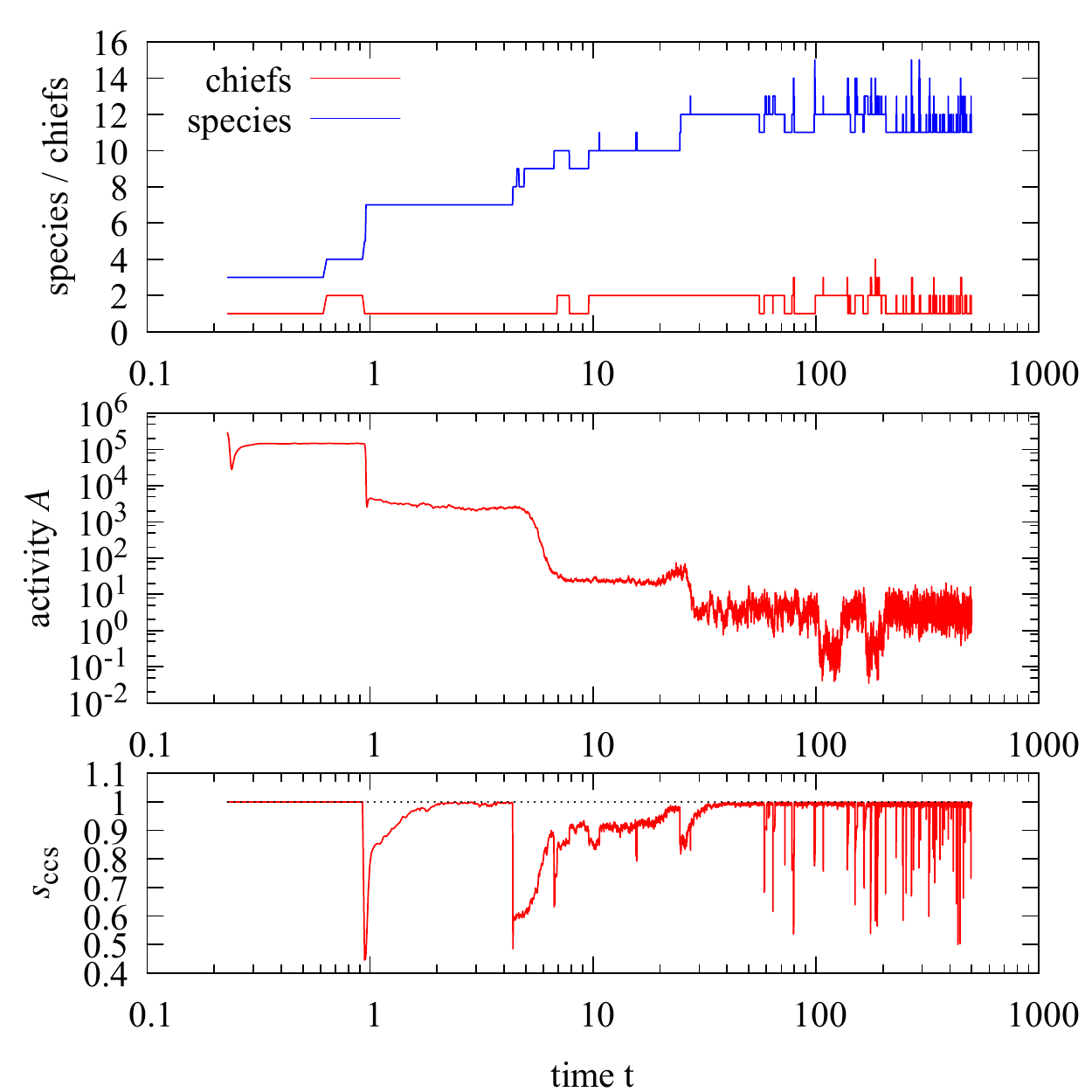}
	\caption{
	Typical development of chief-clan structures (here bootlace) under rare
	random ligation.
	Parameters are chosen as in figure~\ref{fig_transition} with
	$\alpha=10^{-7}$.
	Graphs show the number of species and chiefs (top), an activity that
	measures how ``inverted'' the population is (middle), and the distance
	to the smallest enclosing chief-clan structure (bottom).
	See text for details.
	}
	\label{fig_activity}
\end{figure}
Figure~\ref{fig_activity} shows exemplary results for the formation of a
bootlace structure under the parameters of figure~\ref{fig_transition} with
$\alpha=10^{-7}$.
The stepwise elongation of the bootlace structure is reflected by a steady
increase in species count, but a non-growing number of either one or two chiefs.
While the structure grows, the activity generally decreases,
until it levels off around $t=30$ where the structure is fully established.
Each formation of new chiefs in the structure leads to a temporary decrease
in the otherwise maximal cosine similarity to the corresponding chief-clan
structure.
This decrease in similarity is caused by the temporarily high population count
of the previous chief, which now violates the constant solution.
Similarity is regained with the autocatalytic flow of matter into the newly
formed chief, and is accompanied by a decrease in activity.

These observations indicate that the selection of the bootlace structure
occurs from the onset and is caused by the far-from-equilibrium selection
mechanism that is active during the transient period.
Yet, once the structure is fully developed around $t=30$, the equilibrium
selection mechanism stabilizes the developed population and prevents the
occurence of new chiefs. As a result, fluctuations in species and chief numbers,
activity, and similarity are short-lived.

Visual inspection of several dozens simulation runs confirms this general
behavior independent of the nature of selected structure.
The lower the random ligation rate $\alpha$ the more prominent the
far-from-equilibrium selection. 
For the parameters chosen (c.f. figure~\ref{fig_transition}), clear chief-clan
structures form and remain stable for $\alpha$ between $10^{-7}$ and $10^{-6}$.
For $\alpha=10^{-5}$, the dynamics still yield a transient selection, but
the selected chief-clan structures is no longer maintained in equilibrium,
as random ligations become too common.
Finally, for $\alpha=10^{-4}$ and above, stable chief-clan structures can
no longer form (at least without applying a threshold for the structure
identification).

\section{Model variants}
\label{sec_variants}
We now present several variations of our original reaction system to demonstrate
that the reported selection pressure is structurally robust and can be observed
in similar replicator systems that more closely resemble the chemistry of
nucleic acid replicators.

\subsection{Increased alphabet size}
\begin{figure}
 \begin{center}
  \includegraphics[width=\columnwidth]{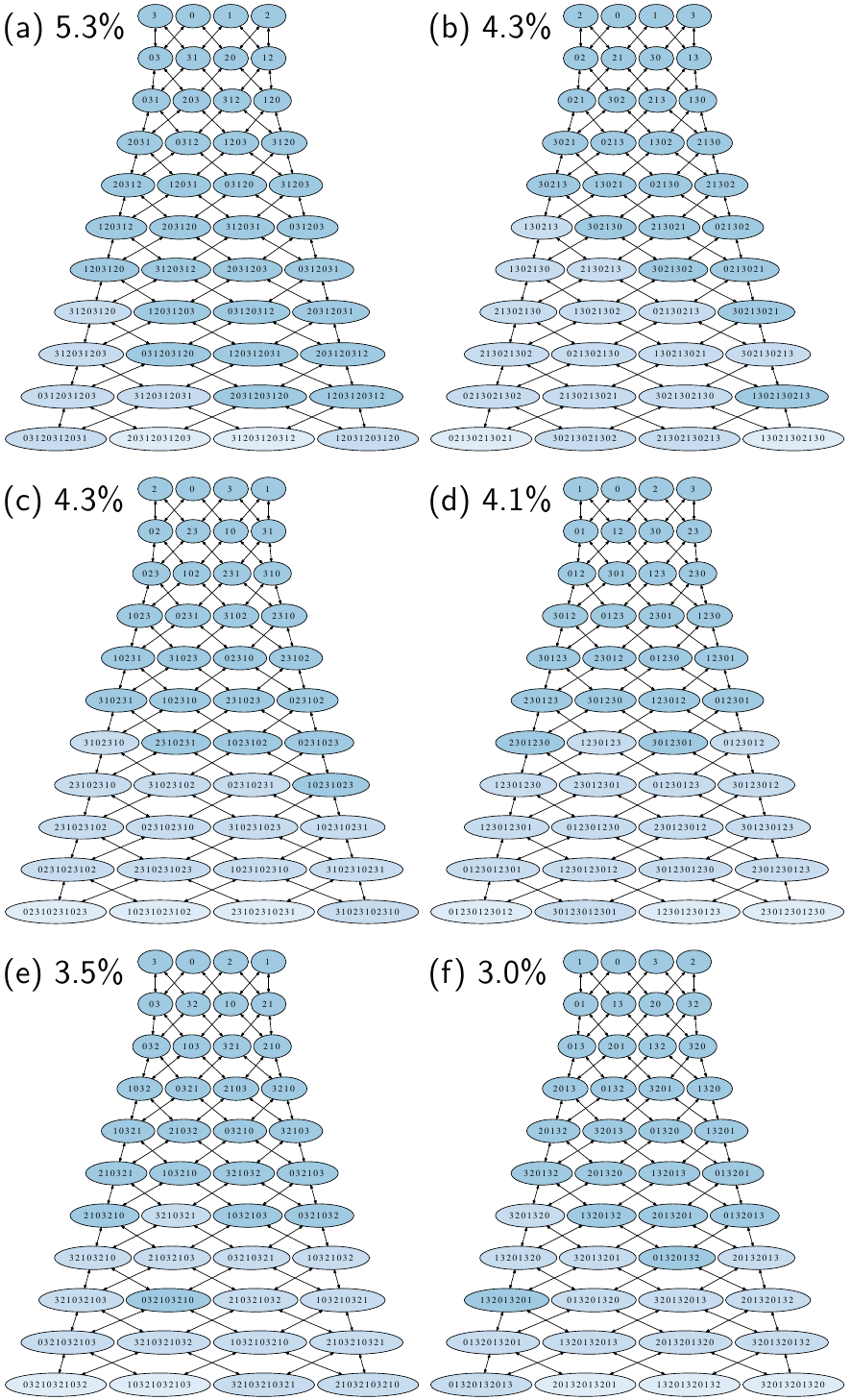}
  \caption{The structures produced from simulations with
  $\mathcal A = \left\{0,1,2,3\right\}$.
  They all belong to the same motif, $\ldots0123\ldots$, and its permutations.
  All six permutations of the motif appear in simulation, occupying in total
  25\% of all populations.
  After the 6th cluster, the structure becomes random.
  The parameters used: $\alpha=10^{-10},\ \beta=10^{-6}$
  with $2\times 10^5$ of 0, 1, 2, and 3.
  The hierarchical clustering was stopped when 30\% of populations are included
  into the largest 10 clusters.
  Species less than 30\% of the constant solution are omitted.
  }
 \end{center}
\label{4mono}
\end{figure}

Alphabet size has a significant impact on the selected population structures.
Here we show populations produced from a pool of four types of monomers
($\mathcal A = \left\{0,1,2,3\right\}$).
Although introducing new monomer types makes the system more complex quickly,
the selection of chief-clan structures does happen as shown in
figure~\ref{4mono}.
The six largest clusters contain about 25\% of all the populations produced.
They all feature permutations of 0123 as motifs.
They are thus a counterpart of the bootlace structure with motif 01.
As in the case of a binary alphabet, symmetry is broken at the level of dimers
and propagates from there through the population structure.
From the seventh cluster onwards [figure~\ref{4mono}(b)], structures lose order
and become more irregular.

\subsection{Mutations}
Evolution arises from variation and selection.
In our basic model, variation is entirely due to random ligation.
It is interesting to see how the reported sequence selection in our model is
affected when potential mutation opens an additional channel for variation.
To this end, point mutations are introduced as a reaction:
\begin{align}
	l'+m'+l.m &\xrightarrow{\;\;c_3|l.m|\;\;} l'.m' + l.m 
\end{align}
where $l'.m'$ is one bit different from $l.m$, and $c_3/c_2=10^{-4}$.

\begin{figure}
\begin{center}
 \includegraphics[width=\columnwidth]{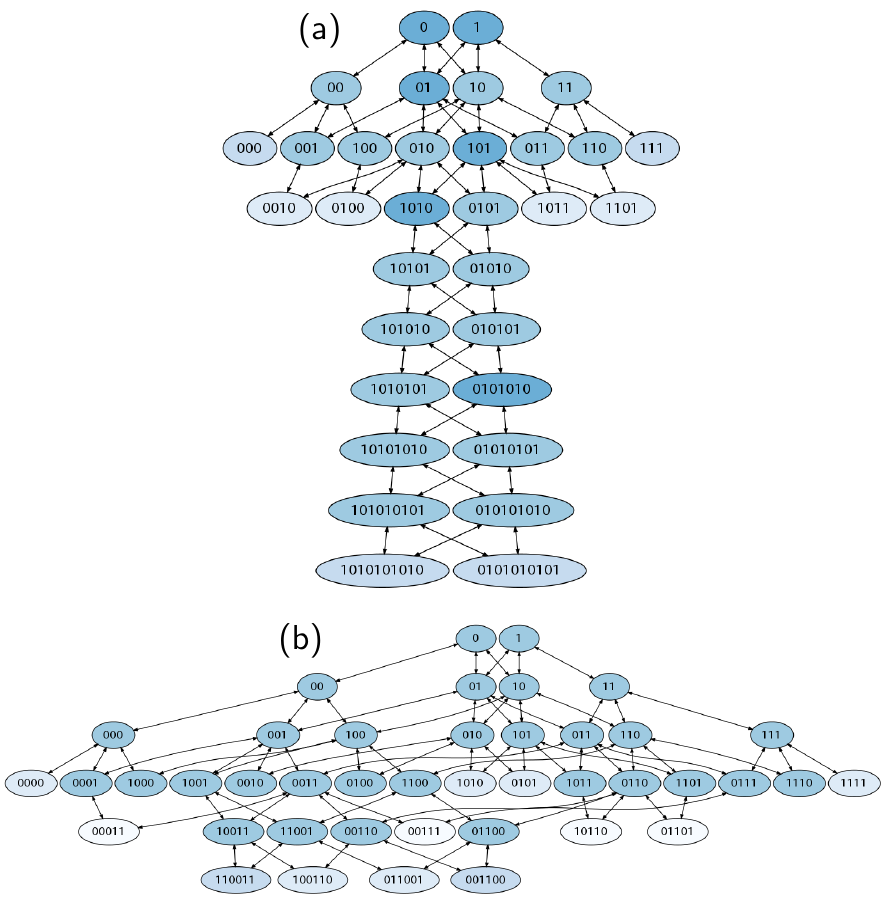}
 \caption{The structures produced from simulations with mutation.
    Bootlace, (a), is more robust against mutation than pinecone1, (b), which is
    disturbed significantly.
    The occupancy is 34\% and 10\% for bootlace and pinecone1, respectively, in
    697 runs.
    The parameters used: $\alpha=10^{-10},\ \beta=10^{-7}$, $c_3=10^{-11}$ with
    $2\times 10^5$ of 0 and 1.
 }
 \label{mutation}
\end{center}
\end{figure}

Figure~\ref{mutation} shows the largest and second largest clusters
found in simulation.  The largest cluster [figure~\ref{mutation}(a)],
occupying 34\% of all populations, is based on the bootlace structure
[figure~\ref{fig_selection}(a)], whereas the second largest cluster
[figure~\ref{mutation}(b)], occupying 10\%, is based on the pinecone1
structure [figure~\ref{fig_selection}(b)], judging from the
tetramer and pentamer sequence motifs.  In figure~\ref{mutation}(a),
despite the fairly large mutation rate, perturbations arising from point
mutations only produce short additional chiefs, and the bootlace
structure remains unperturbed in its longer sequences.  On the other
hand, the distortion in figure~\ref{mutation}(b) is significant.
Actually, all the species shorter than trimers appear.  This indicates
that the robustness against mutations depends on the population
structure, which likely connects to the different stabilities of the
structures.

\subsection{Sequence complementarity}
\begin{figure}
 \begin{center}
 \includegraphics[width=\columnwidth]{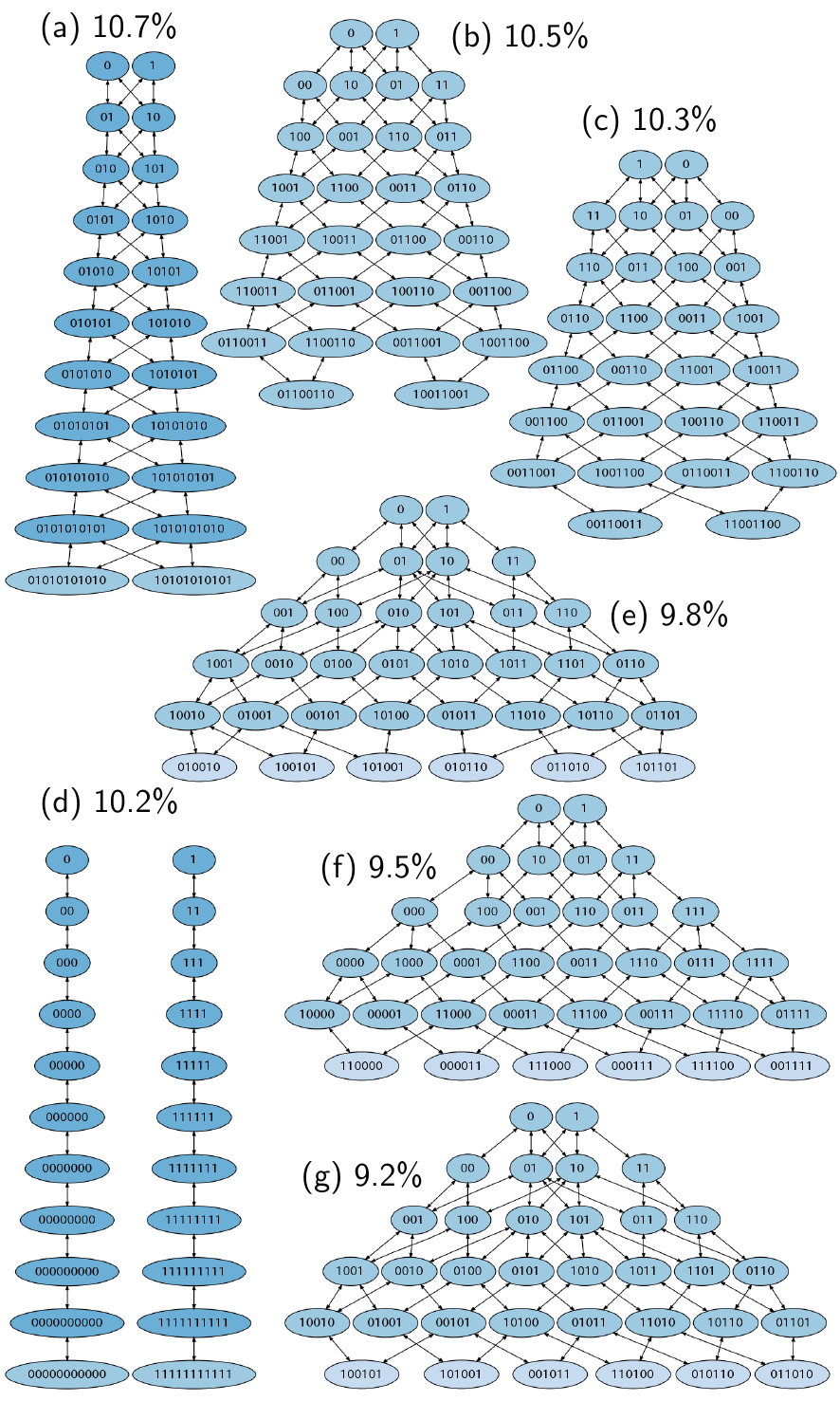}
 \caption{The structures produced by sequence complementary autocatalysis.
 The parameters used: $\alpha=10^{-10},\ \beta=10^{-7}$ with
 $2\times 10^5$ of 0 and 1.
 These seven clusters occupy about 70\% of all the populations produced.
 }
 \label{fig_complement}
 \end{center}
\end{figure}

Sequence complementary hybridization is a characteristic feature of template
directed nucleic acid replication.
We introduce sequence complementarity into our model by defining a
complementarity operator
$\overline{\vphantom{t}\cdot} : \mathcal A^* \rightarrow \mathcal A^*$
recursively as follows:
\begin{align}
	\overline{0} &= 1 \\
	\overline{1} &= 0 \\
	\overline{l.m} &= \overline{l}.\overline{\vphantom{l}m}
\end{align}
This allows us to replace the original template directed ligation
reaction~\eqref{eq_catalyzed} by its sequence-complementary counterpart:
\begin{equation}
	l + m + \overline{l.m} \xrightarrow{\;\;c_2\;\;} l.m + \overline{l.m} .
\end{equation}
Figure~\ref{fig_complement} shows a result of hierarchical clustering out of
1000 runs.
The seven clusters shown in figure~\ref{fig_complement} occupy about 70\% of
all the clusters produced.
This value is slightly smaller than the case in figure~\ref{fig_selection},
but still significant fraction, indicating a strong dynamical selection.
On the other hand, the occupation of each cluster becomes different from the
case in figure~\ref{fig_selection}.
The bootlace structure, for example, is still a biggest cluster, but it
occupies only about 11\% in comparison with 48\% shown in
figure~\ref{fig_selection}.
Pinecone1 [figure~\ref{fig_complement}(b) and (c)] and two-towers
[figure~\ref{fig_complement}(d)] increase their occupancy.
Three new clusters [figure~\ref{fig_complement}(e), (f), and (g)] appear as
replacing the bootlace's occupancy.
One also notices that all the structures shown are symmetric about the center
axis, reflecting the complementarity.

\subsection{Existence of inert food species}
Several previous studies assume the existence of food
species~\cite{hordijk_predicting_2012,derr_prebiotically_2012} that cannot be produced by autocatalysis
and thus have to be present in the system in the beginning or have to be fed
continuously into the system.
In our basic model, monomers are the only food species.
We now analyze a variant where any species up to a certain length $l_{\rm food}$
are food species that cannot undergo autocatalysis.
As a consequence, the system does not have to satisfy the constant
solution~\eqref{constant}.

\begin{figure}
 \begin{center}
 \includegraphics[width=0.7\columnwidth]{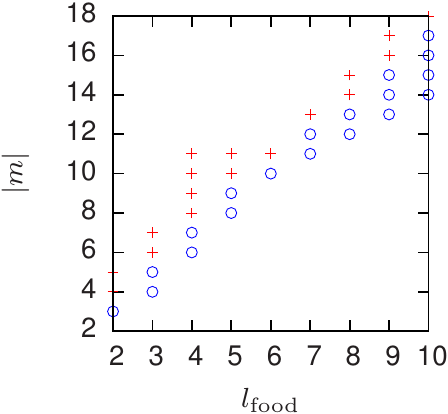}
 \caption{A stability diagram of the bootlace structures with their chief length
    $|m|$ against the maximum length of food species, $l_{\rm food}$.
    Crosses and circles represent where they are stable and unstable,
    respectively.
 }
 \label{food}
 \end{center}
\end{figure}

To examine the effect of food species, we conducted a stability test.
First, a bootlace structure including food species, with two chiefs of length 
$|m|$ is constructed assuming that all species has the same concentration given
by the constant solution.
Then the system is equilibrated.
Figure~\ref{food} shows a diagram indicating whether the bootlace structure is
stable or not, depending on $|m|$ and $l_{\rm food}$.
It can be seen that the structure is stable if the length of the chief is long
enough.
The neccesary length stabilizing the bootlace is roughly linearly proportional
to $l_{\rm food}$.

\begin{figure}
 \begin{center}
 \includegraphics[width=0.7\columnwidth]{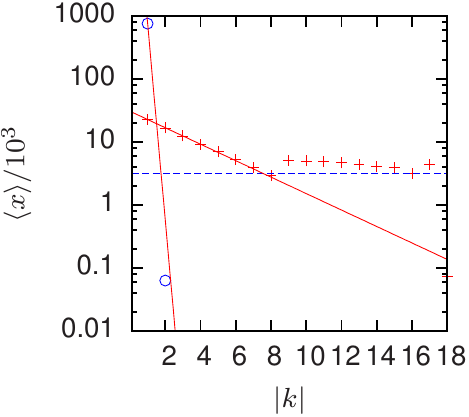}
 \caption{The averaged distribution of species, $\langle x \rangle$, depending
    on their length, $|k|$ when $l_{\rm food}=8$.
    Crosses represent a case starting from a pre-built bootlace structure,
    whereas circles represent a case starting from a pool of monomers.
    Solid lines represent exponential functions.
    Dashed line is the constant solution, $\beta^{-1/2}$.
 }
 \label{food_dist}
 \end{center}
\end{figure}

Figure~\ref{food_dist} shows a distribution of species in equlibrium.
Crosses show a distorted bootlace structure, where the distribution of food
species decreases exponentially.
For autocatalytic species, the constant solution is approximately recovered with
slightly larger values.

If we start simulations from a pool of monomers, monomers must
spontaneously ligate to become longer than $l_{\rm food}$ to initiate
autocatalytic reactions.
It is unlikely then, that autocatalytic species appear spontaneously unless the
concentration of monomers is extremely high.
Circles in figure~\ref{food_dist} show the distribution when the simulation is
started with the same amount of monomers as above, but without a pre-constructed
bootlace structure (crosses in figure~\ref{food_dist}).
Only dimers appear in this case.
Therefore, it is unlikely, if not impossible, for this system to achieve
spontaneously autocatalytic species and the resulting distribution maintained by
autocatalysis.

This barrier between autocatalytic and non-autotacalytic states depends strongly
on the size of food species.
The longer the food species, the higher the barrier.
This observation is important, for example, to a real chemistry or origin-of-life
scenarios.
The available monomer concentration determines how long the food species is
tolerated to achieve an autocatalytically maintained state.

\section{Discussion}

We have presented a simple model of binary self-replicating polymers, in which
exact autocatalysis gives rise to the formation of characteristic chief-clan
structures, where the sequence information of the longest replicators completely
determines the population distribution of the molecular ecosystem.
When starting from a pool of monomers, only few, highly regular sequence
patterns are selected by the dynamics.

We attribute this selection for repetitive sequences to stems in part from
reaction symmetries and in part from a transient concentration of
material in the long sequences. 
Due to the reaction conditions during the transient, most
material is concentrated in the long sequences, the chiefs. In the
transient formation of the longer sequences, mainly from the abundant
chiefs,repetitive sequences both have more and shorter reaction pathways
due to their symmetries. This is true both in the ligation of shorter
sequences into longer sequences and in the breakdown of longer
sequences into shorter sequences. These are the reasons why the
production of repetitive sequences are favoured in such systems. 

Notably, this dynamical selection is intrinsic to our model and does not require
an \emph{a priori} fitness function.
This is important for the study of, for example, prebiotic evolution and
origins of life~\cite{joyce_non-enzyme_1984,breaker_emergence_1994,
vaidya_spontaneous_2012}, as well as protocell research~\cite{
szostack_synthesizing_2001,fellermann_wet_2011,rasmussen_generating_2016},
since it implies that the selection occurs by means of both environmental as
well as internal factors.
This aspect was not elucidated by previous studies where selection was either
induced by means of an external fitness function  or by assigning randomly
attributed catalysis rates.

It is commonly understood that random fluctuations are mostly significant in
systems with small molecular copy numbers and that, in the limit of large
numbers, results of stochastic simulations typically approach expectation values
that can be obtained from equivalent deterministic models.
The situation is radically different in the system presented in this study:
firstly, autocatalytic replication amplifies small random fluctuations in the
transient, which ultimately determines the fate of entire populations on the
system level; and secondly, steady state population numbers are only determined
by reaction constants, but not by the total material concentration.
Increasing the total monomer concentration will not increase the concentration
of molecular species in steady state.
Instead, the system will accommodate for the additional material by generating
larger polymers, up to a length where material becomes sparse and thus
remains susceptible to copy number fluctuations.

Although our exact findings have been derived by analyzing a specific reaction
system (exact autocatalysis), we hypothesize that many of our general results
also hold for similar binary polymer models, for example those employing
sequence dependent replication rates~\cite{derr_prebiotically_2012}, random
cross-catalysis~\cite{langton_spontaneous_1991}, or material flow.

Namely, if studied under conditions where the analysis of our work applies
(sufficiently high molecule numbers, dominant catalysis over random ligation,
and relatively constant environment), any polymer chemistry is expected
to approach some equilibrium distribution.
The phase space profile of these equilibrium states will dictate which
populations are stable and how matter is distributed among their members.
If the kinetics of the polymer chemistry allows for multiple equilibria,
selection is likely to occur.
Either because these are isolated stable states and an initial population has
to commit to one or the other; or because a smaller set of stable states is
selected by noise induced drift, as encountered in the chemistry studied in
this article.
For example, Bagley et al.~\cite{langton_spontaneous_1991} report the
emergence of ``concentration landscapes'' in randomly catalyzing polymer
networks, which we can now identify as equivalents to chief-clan structures.

As long as selected populations introduce novel species that provide resources
and catalysts for higher units of selection (i.e. longer polymers of a
net-catalyzing set), the chemistry meets in principle the conditions for
\emph{continued} selection dynamics during a transient that starts from monomers
or short oligomers.
In the model system discussed in this article, continued selection is
particularly prominent, since every newly created species is an auto-catalyst.
Continued selection is expected to stall once the chemistry does not introduce
novel resources or novel catalysts.
For example, if the chemistry allowed for overhangs around a ligation window,
we expect the symmetry breakdown to stall, once a suitable ligation site has
emerged.

In contrast, we expect that the particular repetitive sequence motifs
that we report here stem from symmetries in the reaction graph of exact
autocatalysis and are likely not prominent in random catalytic chemistries.

The investigation of related chemistries employing sequence dependent and
randomly assigned catalysis is left for future research.

\bibstyle{pre}
\bibliography{references}

\vspace{1em}

\appendix

\section{Fokker Planck derivation}
We study the minimal competing system with four species, where two competing
chiefs $01$ and $10$ compete over resources $0$ and $1$.
Mass conservation maintains the dynamics on a two dimensional manifold,
essentially $x_{01}$ and $x_{10}$ ($x_0 = x_1$ can then be infered).

For brevity, we introduce the notation $X=01, Y=10$, with concentrations
$x=x_{01}, y=x_{10}, z=x+y$, and indistrimnately refer to monomers by $A$
with concentration $a=x_0=x_1$.

Motivated by the observation that there is a line of steady state solutions
(any combination of $x$ and $y$ that fulfills $a = \beta^{1/2}$)
that are all of marginal stability (zero eigenvalues), we introduce a
simplification that constrains the system to maintain equilibrium by only
distributing material among chiefs.

To achieve this, we construct a constrained stochastic process, where no two
degradations nor two ligations can follow each other without an intermediate
event of the other type.
Each net degradation-ligation event thus maintains the initial distance from the
equilibrum line.

Technically, we introduce a stochastic net process with two Markovian
degradation reactions, and two delayed time ligation reactions whose
rates $\beta$ are functions of the time $t-t'$ since the last degradation event.
With
\begin{equation}
    \int_{t'}^\infty \beta(t-t') dt = \beta ,
\end{equation}
we ensure that the original ligation rate is maintained in the long term.
Informally, the constrained process can be written as
\begin{align}
    X  &\xrightarrow{1} 2 A \xrightarrow{\beta(t-t') x} X \label{eq_combined_1} \\
    X  &\xrightarrow{1} 2 A \xrightarrow{\beta(t-t') y} Y \label{eq_combined_2} \\
    Y  &\xrightarrow{1} 2 A \xrightarrow{\beta(t-t') x} X \\
    Y  &\xrightarrow{1} 2 A \xrightarrow{\beta(t-t') y} Y \label{eq_combined_4} .
\end{align}
We now let
\begin{equation}
    \beta(t-'t) \rightarrow \beta\delta_{t-t'}
\end{equation}
approach a Dirac distribution, so that ligation follows degradation
\emph{immediately}.

Degradation events occur as previously with rate constant 1, but only a
certain fraction (e.g. $y/(x+y)$ for equation~\eqref{eq_combined_2}) lead
to an actual change in the system state.
The net conversion rates are therefore the degradation rate (having been set
to one per unit time in the dimensionless system) times the probability to lead
to a system change.
The net process therefore reads:
\begin{gather}
    X \xrightarrow{(z-x)/z} Y \label{eq_reduced_Y} \\
    Y \xrightarrow{x/z} X     \label{eq_reduced_X}
\end{gather}
Note that the total propensity of this net process, $(z-x)/z + x/z = z$,
equals the total propensity of degradations in the original process.

\begin{figure}
    \includegraphics[width=\columnwidth]{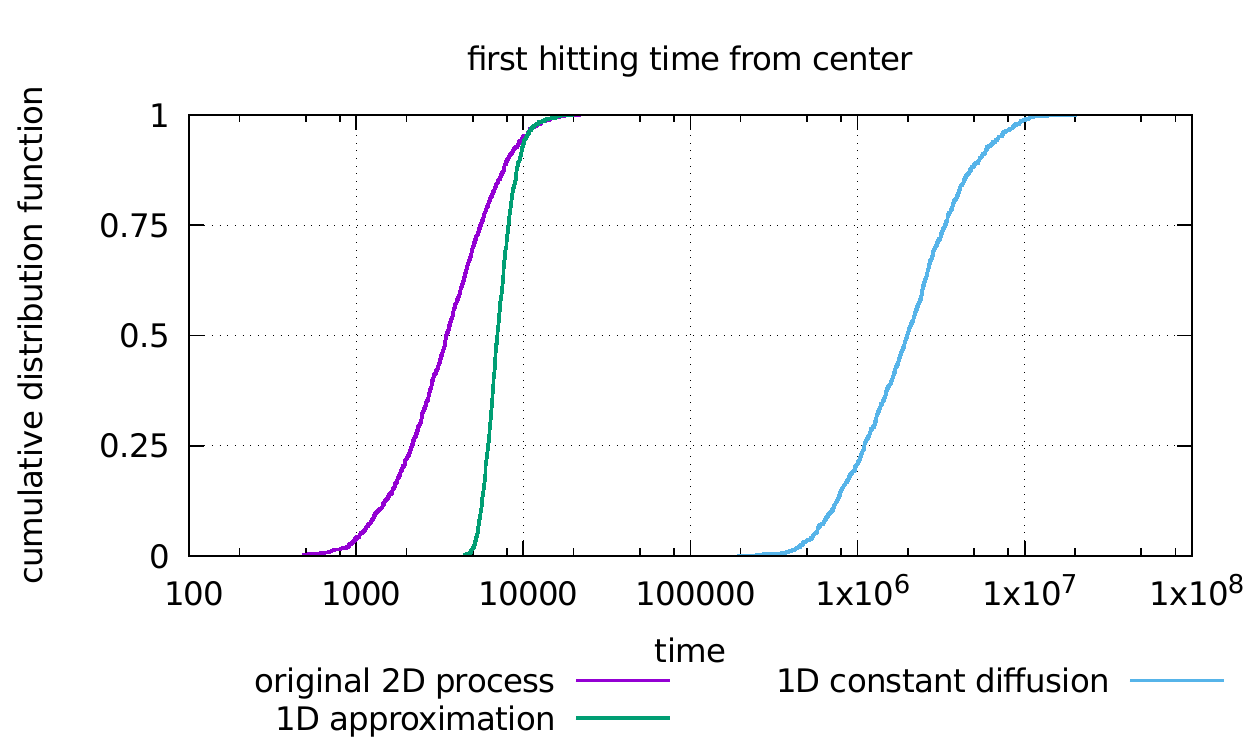}
    \caption{
    Cumulative hitting time distributions of different random walks starting
    in the inital state $x(0)=\frac z 2=\frac {x^*} 2$ for $z=\sqrt{10^{7}}$.
    Distribution functions have been obtained from 1000 numerical samples
    for each of \emph{purple:} the original 2D process defined by equations (1)
    and (3), \emph{green:} our 1D approximation of equation
    (\ref{eq_reduced_Y}), and \emph{blue:} a random walk with constant
    diffusion for reference.
    }
    \label{fig_hitting_times}
\end{figure}
To compare the quality of this approximation, we sampled 1000 instantiations
of the original 2D process and of our 1D simplification and compare the
cumulative probability distributions of first hitting times for walks that
start at $x(0)=\frac z 2$ and reach one of the absorbing bundaries $0$ or $z$.
Results are shown in figure~\ref{fig_hitting_times}.

The results reveal that the 1D approximation generates hitting time distributions
with the same general shape than the original process, but overestimates
typical hitting times -- that is, the unconstrained process typically terminates
\emph{faster} than the constrained process.
The approximation is more accurate for late hitting times, and approaches the
original hitting time distrubtion.

Figure~\ref{fig_hitting_times} also reveals that both the original and the
constrained process complete orders of magnitude faster than a random walk
where $X$ transforms into $Y$ and vice versa with constant rate $z$.

\begin{widetext}
We now derive the Master and Fokker Planck equations for the stochastic
process~\eqref{eq_reduced_Y}---\eqref{eq_reduced_X}.
Since mass is conserved in this system, this corresponds to a one dimensional
random walk, here expressed for the species X=01:
\begin{equation}
	\ldots \ce{<=>} x-1 \ce{<=>[$(x-1)(z-x+1)/z$][$x(z-x)/z$]} x \ce{<=>[$x(z-x)/z$][$(x+1)(z-x-1)/z$]} x+1 \ce{<=>} \ldots
\end{equation}
The stochastic process has the associated Master equation
\begin{align}
	\frac{\partial}{\partial\mathsf t} P(x,t)
	&= - 2\frac{x(z-x)}z P(x,t) + \frac{(x-1)(z-x+1)}z P(x-1,t) + \frac{(x+1)(z-x-1)}z P(x+1,t) \\
	&= - 2\frac{x(z-x)}z P(x,t) \nonumber\\
	&\qquad\qquad + \frac{x(z-x)}z P(x-1,t) + \frac{x}z P(x-1,t) - \frac{z-x}z P(x-1,t) - \frac{1}z P(x-1,t) \nonumber\\
	&\qquad\qquad + \frac{x(z-x)}z P(x+1,t) - \frac{x}z P(x+1,t) + \frac{z-x}z P(x+1,t) - \frac{1}z P(x+1,t) \\
	&=  \frac{x(z-x)}z \left[ P(x+1,t) - 2P(x,t) + P(x-1,t) \right] \nonumber\\
	&\qquad\qquad + \frac{z-2x}z \left[ P(x+1,t) - P(x-1,t) \right] - \frac{1}z P(x+1,t) - \frac{1}z P(x-1,t) \\
	&=  \frac{x(z-x)-1}z \left[ P(x+1,t) - 2P(x,t) + P(x-1,t) \right] \nonumber \\
	&\qquad\qquad + \frac{z-2x}z \left[ P(x+1,t) - P(x-1,t) \right] - \frac{2}z P(x,t) .
\end{align}
Approximating
\begin{align}
	P(x+1,t) - P(x-1,t) &\approx 2 \frac{\partial}{\partial\mathsf x} P(x,t) \\
	P(x+1,t) - 2P(x,t) + P(x-1,t) &\approx \frac{\partial^2}{\partial\mathsf x^2} P(x,t) ,
\end{align}
we derive a Fokker Planck equation with zero drift and non-constant diffusion:
\begin{align}
	\frac{\partial}{\partial\mathsf t} P(x,t)
	&\approx  \frac{x(z-x)-1}z \frac{\partial^2}{\partial\mathsf x^2} P(x,t) + 2\frac{z-2x}z \frac{\partial}{\partial\mathsf x} P(x,t) - \frac{2}z P(x,t) \\
	&= \frac{z-2x}z \frac{\partial}{\partial\mathsf x} P(x,t) + \frac{x(z-x)-1}z \frac{\partial^2}{\partial\mathsf x^2} P(x,t) + \frac{z-2x}z \frac{\partial}{\partial\mathsf x} P(x,t) - \frac{2}z P(x,t) \\
	&= \frac{\partial}{\partial\mathsf x} \frac{x(z-x)-1}z \frac{\partial}{\partial\mathsf x} P(x,t) + \frac{x(z-x)-1}z \frac{\partial^2}{\partial\mathsf x^2} P(x,t) +\frac{\partial}{\partial\mathsf x} \frac{z-2x}z P(x,t) + \frac{z-2x}z \frac{\partial}{\partial\mathsf x} P(x,t) \\
	&= \frac{\partial}{\partial\mathsf x}\left[ \frac{x(z-x)-1}z \frac{\partial}{\partial\mathsf x}P(x,t) \right] + \frac{\partial}{\partial\mathsf x} \left[ \frac{z-2x}z P(x,t) \right] \\
	&= \frac{\partial}{\partial\mathsf x}\left[ \frac{x(z-x)-1}z \frac{\partial}{\partial\mathsf x}P(x,t) + \frac{\partial}{\partial\mathsf x} \frac{x(z-x)-1}z P(x,t) \right] \\
	&= \frac{\partial^2}{\partial\mathsf x^2}\left[ \frac{x(z-x)-1}z  P(x,t) \right] \\
	&= \frac{\partial^2}{\partial\mathsf x^2}\left[ D(x)  P(x,t) \right]
\end{align}
with non-constant diffusion $D(x)=\frac{x(z-x)-1}z$.
Note that diffusion is maximal for $x=z/2$, i.e. when both chiefs are equally
populated, and becomes negative for $x=0$ and $x=z$, which reflects the fact that
these are absorbing boundary conditions.

\end{widetext}

\end{document}